\documentclass[aps,pra,twocolumn,superscriptaddress,longbibliography,10pt]{revtex4-2}

\usepackage{amsmath,amssymb,bm}
\usepackage{graphicx}
\usepackage{xcolor}
\usepackage{microtype}
\usepackage{hyperref}
\usepackage{tikz}
\usepackage{float}
\usetikzlibrary{arrows.meta,positioning,calc,fit,backgrounds}
\hypersetup{colorlinks=true,linkcolor=blue!60!black,citecolor=blue!60!black,urlcolor=blue!60!black}

\newcommand{\Fmarg}{F_{\mathrm{marg}}}
\newcommand{\Fcond}{F_{\mathrm{cond}}}
\newcommand{\Fjoint}{F_{\mathrm{joint}}}
\newcommand{\FQ}{F_{Q}}
\newcommand{\Eb}{\mathbb{E}_{b}}
\newcommand{\Es}{\mathbb{E}_{s}}
\newcommand{\that}{\hat{\theta}}
\newcommand{\tstar}{\theta^{\ast}}
\newcommand{\khat}{\hat{\kappa}}
\newcommand{\Etw}{\mathcal{E}_{\mathrm{tw}}}

\begin{document}

\title{Recovering Coherent Errors from Randomized-Compiling Labels}

\author{Kushagra Vyas}
\email{ph25mtech14002@iith.ac.in}
\affiliation{Department of Physics, Indian Institute of Technology
Hyderabad, Sangareddy 502284, India}

\date{\today}

\begin{abstract}
Randomized compiling converts a coherent gate error into a stochastic
one by dressing each gate with a randomly chosen Pauli operator, and
the standard view is that the coherent information is lost once
outcomes are averaged over these random choices. This is incorrect:
an exact Fisher-information identity shows that the averaged data and
the correlation between the outcome and the random choice, the twirl
label, together conserve the full information, and only the label is
normally discarded. Retaining it turns an unmodified
randomized-compiling run into an unbiased estimator for the coherent
error, at no added circuit cost, valid under realistic decoherence and
at any circuit depth. The same construction extends to coupling
between qubits through two further mechanisms: a neighboring qubit
prepared in a known, randomized state probes static coupling, and
randomizing the pulse count used to prepare that state isolates a
second, physically distinct leakage effect found unexpectedly on real
hardware during this work. All three results are derived exactly,
checked against closed-form oracles, and tested on a 127-qubit IBM
processor. The single-qubit case is confirmed cleanly: an injected
phase is recovered to 0.006~rad while the same data without labels
shows no signal. The two-qubit extensions are validated at the level
of the recovery circuit, with clean positive controls on hardware,
though the coupling strengths themselves remain below the resolution
of the shot budgets used here.
\end{abstract}

\maketitle

\section{Introduction}

A miscalibrated gate is typically wrong in the same way every time it
is applied. This repeatability is what makes coherent errors, small
systematic over- or under-rotations arising from calibration drift or
unwanted coupling, more damaging than ordinary stochastic noise.
As identical errors add up coherently in amplitude rather than
incoherently in probability, their contribution to a circuit's total
error grows quadratically with depth rather than linearly~\cite{Wallman2016}.

Randomized compiling addresses this by breaking the repetition. The
compiler inserts a Pauli operator drawn at random before and after
each gate, chosen so that the pair leaves the intended operation
unchanged~\cite{Wallman2016,Hashim2021}. A coherent error located
between the two insertions does not cancel in the same way; instead it
is conjugated by the inserted Pauli, and for a Pauli-type generator
this simply flips the sign of the error whenever the inserted operator
anticommutes with it. Averaged over many shots, the error therefore
appears with either sign roughly half the time, and what remains in
the pooled measurement statistics resembles a Pauli channel rather
than a fixed, systematic bias. This is a real and useful effect, and
it is the reason the technique has become standard practice.

The usual description of randomized compiling stops at this averaging
step. Once the shots are pooled together, the first-order dependence
on the coherent angle cancels along with the sign, so the pooled
distribution carries little information about the original error. The
natural conclusion, stated in various forms throughout the literature
on randomized compiling and cycle benchmarking, is that the coherent
parameters have effectively been converted into noise and can no
longer be recovered separately.

This conclusion is not correct. The sign that flips the coherent error
on a given shot is determined entirely by the classical control
software, which selects the twirling Pauli before the circuit runs and
therefore knows the resulting sign exactly. Conditioned on that sign,
the coherent error produces a fixed and ordinary shift in the outcome
statistics for that shot. Pooling the data across shots discards the correlation between the recorded sign and the measured
outcome, and it is precisely that correlation which carries the
information the averaged data lack. Retaining the sign, which we refer
to as the twirl label, and correlating it against the outcome yields
an estimator for the coherent error that costs nothing beyond the
randomized-compiling run that was going to be performed anyway.

Section~\ref{sec:conservation} derives this result formally: the
Fisher information about a coherent angle splits into a component
retained by the averaged data and a positive-semidefinite component
carried by the label-outcome correlation, with their sum fixed
independently of the twirl distribution. Section~\ref{sec:noise}
extends the analysis to realistic decoherence and gives an estimator
that remains unbiased once a single visibility factor, already
available from ordinary cycle benchmarking, is divided out.
Section~\ref{sec:depth} considers circuit depth and shows that
retaining labels eventually outperforms not twirling at all, a result
that is not obvious in advance. Sections~\ref{sec:zz}
and~\ref{sec:drive} extend the construction to coupling between
qubits. We present these results in the order in which they were
obtained: a static-coupling theorem that a subsequent hardware run
could not confirm above its noise floor, followed by an unanticipated
leakage mechanism discovered while investigating that null result and
requiring a separate derivation. Sections~\ref{sec:sim}
through~\ref{sec:hw-drive} report the corresponding simulation and
hardware results, and Sec.~\ref{sec:discussion} compares the overall
approach with gate set tomography~\cite{Nielsen2021}, hidden
inverses~\cite{Zhang2022}, and pseudo twirling~\cite{Santos2024}, the
closest prior work we are aware of and one built on a different
mechanism.

\begin{figure*}[t]
\centering
\resizebox{\textwidth}{!}{%
\begin{tikzpicture}[
  x=1mm,y=1mm,
  font=\scriptsize,
  panel/.style   ={draw=black!45,fill=white,rounded corners=1mm,line width=0.4pt},
  hdr/.style     ={draw=black!55,fill=black!10,rounded corners=1mm,
                   text width=42mm,align=center,inner sep=1.5mm,
                   font=\bfseries\scriptsize},
  bx/.style      ={draw=black!40,fill=white,rounded corners=1mm,
                   text width=42mm,align=center,inner sep=1.5mm,line width=0.4pt,
                   execute at begin node={\hyphenpenalty=10000\exhyphenpenalty=10000}},
  key/.style     ={bx,draw=orange!85!black,fill=orange!8,line width=0.7pt},
  res/.style     ={bx,draw=blue!55!black,fill=blue!5,line width=0.6pt},
  gate/.style    ={draw=black!70,fill=black!4,rounded corners=0.3mm,line width=0.4pt,
                   minimum width=5.4mm,minimum height=5.4mm,inner sep=0pt,font=\scriptsize},
  pauli/.style   ={gate,fill=blue!10,draw=blue!55!black},
  flow/.style    ={-{Stealth[length=2.4mm,width=1.8mm]},line width=0.7pt,black!65},
  wire/.style    ={line width=0.4pt,black!75}
]

\node[hdr] (h1) at (25 ,0) {STAGE 1\\[0.3mm] Twirled circuit and the label $s$};
\node[hdr] (h2) at (78 ,0) {STAGE 2\\[0.3mm] Keep the label of every shot};
\node[hdr] (h3) at (131,0) {STAGE 3\\[0.3mm] Unbiased estimate of $\theta$};
\node[hdr] (h4) at (184,0) {STAGE 4\\[0.3mm] Recalibrate, and reach further};

\node[panel,minimum width=45mm,minimum height=35mm,anchor=north] (circ) at (25,-8) {};
\node[anchor=north,align=center,text width=42mm,font=\tiny] at (25,-9.0)
      {one dressed cycle of an $n$-qubit circuit};

\draw[dashed,draw=orange!85!black,line width=0.6pt,rounded corners=0.8mm]
      (10.5,-14.5) rectangle (33.5,-37.5);
\node[anchor=north,font=\scriptsize,text=orange!70!black] at (22,-37.6)
      {randomized--compiling layer};

\foreach \yy in {-19,-26,-34}{\draw[wire] (8,\yy) -- (36.5,\yy);}
\node[font=\scriptsize] at (5.6,-19) {$|{+}\rangle$};
\node[font=\scriptsize] at (5.6,-26) {$|0\rangle$};
\node[font=\scriptsize] at (5.6,-34) {$|0\rangle$};
\node[font=\tiny] at (5.6,-30) {$\vdots$};
\node[font=\tiny] at (22,-29.8) {$\vdots$};

\foreach \yy in {-19,-26,-34}{
  \node[pauli] at (14,\yy) {$P$};
  \node[pauli] at (30,\yy) {$P$};
}
\draw[gate] (19.2,-16.3) rectangle (24.8,-28.7);
\node[font=\scriptsize] at (22,-22.5) {$G$};
\node[gate] at (22,-34) {$G'$};
\foreach \yy in {-19,-26,-34}{
  \draw[panel] (36.5,\yy-2.7) rectangle (41.9,\yy+2.7);
  \draw[black!70,line width=0.4pt] (37.7,\yy-1.2) arc (180:0:1.5);
  \draw[black!70,line width=0.4pt,-{Stealth[length=1mm]}] (39.2,\yy-1.2) -- (40.4,\yy+1);
}
\node[font=\tiny,anchor=west] at (42.6,-26) {$b_m$};

\node[key,below=2.2mm of circ,anchor=north] (conj)
 {$P\,e^{-i\theta_a G_a/2}\,P=e^{-is_a\theta_a G_a/2}$\\[0.6mm]
  \textbf{twirl label} $s_a=+1$ ($P$ commutes with $G_a$),\\
  $s_a=-1$ ($P$ anticommutes)};

\node[bx,below=2.2mm of conj,anchor=north] (soft)
 {The compiler draws $\bm s\sim\pi(\bm s)$ \emph{before} execution, so
  the classical control software already knows every sign exactly, and
  $\pi(\bm s)$ carries no dependence on $\bm\theta$.};

\node[bx,anchor=north] (shots) at (78,-8)
 {\textbf{Record the pair $(b_m,s_m)$ per shot}\\[0.8mm]
  \renewcommand{\arraystretch}{1.05}
  \begin{tabular}{@{}ccc@{}}
   shot & outcome & label\\[0.2mm]\hline
   $1$      & $b_1$      & $s_1$\\
   $\vdots$ & $\vdots$   & $\vdots$\\
   $M$      & $b_M$      & $s_M$\\
  \end{tabular}\\[1mm]
  $P(b\mid s,\theta)=\tfrac12\bigl(1+bs\sin\theta\bigr)$};

\node[bx,below=2.2mm of shots,anchor=north] (pooled)
 {\textbf{Standard RC: pool the shots}\\[0.5mm]
  $\bar P(b\mid\bm\theta)=\sum_{\bm s}\pi(\bm s)P(b\mid\bm s,\bm\theta)$\\[0.6mm]
  $\Fmarg=(1-T)^2F_0\;\xrightarrow[\text{full twirl}]{}\;0$};

\node[key,below=2.2mm of pooled,anchor=north] (corr)
 {\textbf{This work: correlate label with outcome}\\[0.5mm]
  $\mathbb{E}[sb]\simeq\dfrac{1}{M}\displaystyle\sum_{m=1}^{M}s_mb_m$\\[0.6mm]
  $\Delta=(2T-T^2)F_0\;\xrightarrow[\text{full twirl}]{}\;F_0$};

\node[res,below=2.2mm of corr,anchor=north] (cons)
 {\textbf{Conservation law}\\[0.5mm]
  $\Fcond=\Fmarg+\Delta\;\succeq\;\Fmarg$\\[0.6mm]
  The twirl only \emph{moves} the coherent information; the sum is
  fixed by the circuit and the noise.};

\node[bx,anchor=north] (vis) at (131,-8)
 {\textbf{Visibility, from the same run}\\[0.5mm]
  $\Etw^{\dagger}(Y)=t_YI+\lambda_YY,\qquad v=\lambda_Y$\\[0.6mm]
  a Pauli fidelity on the diagonal of the twirled transfer matrix,
  already produced by ordinary cycle benchmarking.};

\node[bx,below=2.2mm of vis,anchor=north] (mean)
 {\textbf{Noisy conditional mean}\\[0.5mm]
  $m(s)=v\,s\sin\theta+o$\\[0.6mm]
  Correlating against the label removes the offset $o$ automatically at
  full twirl: $\;\mathbb{E}[sb]=v\sin\theta$.};

\node[key,below=2.2mm of mean,anchor=north] (est)
 {\textbf{Unbiased estimator}\\[0.8mm]
  $\displaystyle \that=\arcsin\!\left(\frac{\mathbb{E}[sb]}{v}\right)=\theta$};

\node[res,below=2.2mm of est,anchor=north] (depth)
 {\textbf{And it survives depth}\\[0.5mm]
  $\displaystyle \mathbb{E}_{s}[F_Q]=\sum_{\ell=1}^{L}\sin^{2}(\ell\alpha)\;\longrightarrow\;\frac{L}{2}$\\[0.6mm]
  unbounded in $L$, whereas the untwirled circuit saturates at
  $1/\sin^{2}(\alpha/2)$.};

\node[key,anchor=north] (recal) at (184,-8)
 {\textbf{Feed $\that$ back into the compiler}\\[0.8mm]
  \begin{tikzpicture}[x=1mm,y=1mm,baseline]
    \node[gate,fill=orange!12,draw=orange!85!black,anchor=center,inner sep=2pt] (rz) at (8,0) {$R_z$};
    \draw[wire] (0,0) -- (rz.west);
    \draw[wire] (rz.east) -- (16,0);
  \end{tikzpicture}\\[0.5mm]
  $\theta_{\mathrm{new}}\simeq\theta-\that$\\[0.4mm]
  a virtual frame update, so the drifting coherent phase is cancelled
  on the next run at zero pulse cost.};

\node[bx,below=2.2mm of recal,anchor=north] (multi)
 {\textbf{More labels, more physics}\\[0.5mm]
  Label the neighbour's prepared state ($c_j$) and its pulse count ($p_j$):\\[0.6mm]
  $\theta_1(c_j,p_j)=\Theta+Kc_j-\Phi p_j$\\[0.6mm]
  which separates the qubit's own error $\theta_1^{(0)}$, the static
  $ZZ$ coupling $\kappa_{1j}$, and the per-pulse leakage $\phi_{1j}$.};

\node[res,below=2.2mm of multi,anchor=north] (hw)
 {\textbf{On hardware} (\texttt{ibm\_marrakesh}, $L\le32$)\\[0.5mm]
  $|\that-\tstar|\le0.006$~rad across the injected sweep, while the
  identical shots analysed with the labels discarded show no signal at
  any depth.};

\draw[flow] (48.5,-40) -- (55.5,-40);
\draw[flow] (101.5,-40) -- (108.5,-40);
\draw[flow] (154.5,-40) -- (161.5,-40);

\draw[flow,dashed]
   (recal.east) -- (211,0 |- recal.east) -- (211,9) -- (-6,9)
   -- (-6,0 |- circ.west) -- (circ.west);
\node[anchor=south,font=\scriptsize,text=black!65] at (102,9.6)
   {re-run the same randomized-compiling circuits with the updated calibration};

\end{tikzpicture}}
\caption{\textbf{Recovering coherent errors from randomized-compiling
labels.} Standard randomized compiling dresses each gate with a random
Pauli, which conjugates a coherent error into a sign-flipped copy of
itself (stage~1) and then averages the sign away (stage~2, upper
branch). The sign is not random to the experiment: the classical
control software chooses it, so it can be stored with the shot. The
correlation between the stored label and the measured outcome
(stage~2, lower branch) carries exactly the Fisher information the
average discards, and dividing out a visibility already available from
the same run gives an unbiased estimate of the coherent angle
(stage~3) that grows more informative with circuit depth. The
recovered angle can be returned to the compiler as a virtual-$Z$
correction, and the same construction extends to inter-qubit coupling
once additional labels are retained (stage~4). No circuit is added
anywhere in the loop.}
\label{fig:workflow}
\end{figure*}
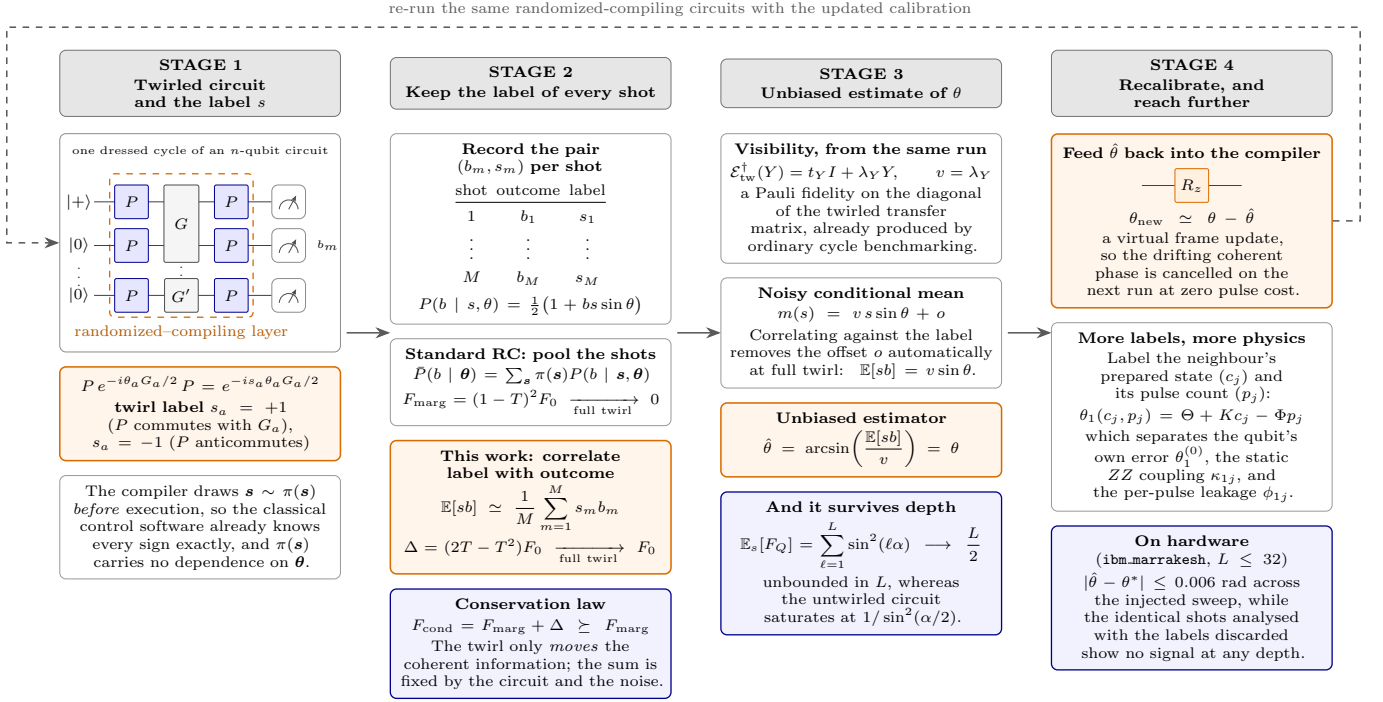

\section{Fisher information conservation}
\label{sec:conservation}

\subsection{Setup}

Write the coherent errors present in a circuit as
$\bm\theta=(\theta_1,\ldots,\theta_K)$, one angle per gate location and
error type, with generator $G_a$ (a Pauli or Pauli string) at each
location. Twirling dresses location $a$ with a Pauli $P$ satisfying
$P^2=I$, leaving the intended gate untouched while conjugating the
error:
\begin{equation}
P\,e^{-i\theta_a G_a/2}\,P = e^{-is_a\theta_a G_a/2},
\label{eq:conj}
\end{equation}
with $s_a=+1$ when $P$ commutes with $G_a$ and $s_a=-1$ when it
anticommutes. The realized sign vector on one shot,
$\bm s=(s_1,\ldots,s_K)\in\{\pm1\}^K$, is the twirl label. It is drawn
from a distribution $\pi(\bm s)$ fixed entirely by the compiler,
independent of $\bm\theta$, since the randomness is classical and
known before the circuit is submitted for execution. A shot returns
outcome $b$ with probability $P(b\mid\bm s,\bm\theta)$; standard
practice keeps only the average
\begin{equation}
\bar P(b\mid\bm\theta) = \sum_{\bm s}\pi(\bm s)\,P(b\mid\bm s,\bm\theta),
\label{eq:marginal}
\end{equation}
which is what makes the effective channel a Pauli channel. The label
itself is discarded once the shot has been recorded.

Three Fisher information matrices describe what different amounts of
bookkeeping retain: $\Fjoint$ for the joint distribution of $(b,\bm
s)$; $\Fcond=\Es[\mathcal I(\bm\theta\mid\bm s)]$, the label-averaged
conditional information; and $\Fmarg$, the Fisher information of
Eq.~\eqref{eq:marginal}.

\subsection{The conservation law}

\emph{Score identity.} Since $\pi(\bm s)$ carries no $\bm\theta$
dependence, the joint log-likelihood's derivative reduces to the
conditional one,
$\partial_a\ln P(b,\bm s\mid\bm\theta)=\partial_a\ln P(b\mid\bm
s,\bm\theta)$, giving
\begin{equation}
\Fjoint(\bm\theta) = \Fcond(\bm\theta).
\label{eq:score}
\end{equation}
Knowing which Paulis were drawn tells you nothing about $\bm\theta$ on
its own; Eq.~\eqref{eq:score} states that the joint observation is
worth exactly the label-averaged conditional experiment.

\emph{Chain rule.} Factor the same joint distribution the other way,
$P(b,\bm s\mid\bm\theta)=\bar P(b\mid\bm\theta)P(\bm s\mid
b,\bm\theta)$. The cross term in the resulting Fisher matrix vanishes
because, at fixed $b$, $\mathbb E_{\bm s\mid
b}[\partial_a\ln P(\bm s\mid b,\bm\theta)]=\partial_a\sum_{\bm
s}P(\bm s\mid b,\bm\theta)=0$. What remains is
\begin{equation}
\Fjoint = \Fmarg + \Delta,\qquad \Delta\equiv\Eb[F_{\bm s\mid b}]\succeq0,
\label{eq:chain}
\end{equation}
with $\Delta$ positive semidefinite because each $F_{\bm s\mid b}$ is
an averaged outer product of scores.

\emph{Conservation.} Combining Eqs.~\eqref{eq:score}
and~\eqref{eq:chain},
\begin{equation}
\boxed{\Fcond = \Fmarg + \Delta\;\succeq\;\Fmarg.}
\label{eq:conserve}
\end{equation}
The left side is fixed by the circuit and the noise, while the twirl
distribution can only move information between the two terms on the
right without changing their sum. Standard randomized compiling
reports $\Fmarg$ and discards $\Delta$.

\subsection{The trade explicitly}

Consider the simplest case: a $Z$-type error of angle $\theta$ acting
on a qubit prepared in $|+\rangle$, with sign label $s$ and a
measurement in the $Y$ basis. The outcome probability is
$P(b\mid s,\theta)=\tfrac12(1+bs\sin\theta)$, and at fixed $s$ the
Fisher information evaluates to $F_0=1$ for either sign, equal to the
quantum Fisher information for phase encoding with generator
$Z/2$~\cite{Braunstein1994}. The conditional experiment therefore
already saturates the quantum limit. Averaging over the label replaces
$s$ with its mean $\mu=\mathbb E[s]$, and writing the tailoring
strength as $T=1-|\mu|$,
\begin{equation}
\underbrace{\Fmarg}_{\text{kept}} = (1-T)^2F_0,\qquad
\underbrace{\Delta}_{\text{discarded}} = (2T-T^2)F_0,
\label{eq:complementarity}
\end{equation}
to leading order in $\theta$, with $\Fmarg+\Delta=F_0$ exactly. At full
twirl the averaged data contain nothing about $\theta$, yet $\Delta$
holds the entire quantum-limited signal. Nothing has actually been
destroyed; it has only been moved somewhere that standard practice
does not look.

\section{Recovery under real noise}
\label{sec:noise}

Real qubits decay and dephase, and it matters how this interacts with
the label correlation. Represent a single-qubit channel by its Pauli
transfer matrix
$\Lambda=\left(\begin{smallmatrix}1&\mathbf 0^\top\\
\mathbf t&A\end{smallmatrix}\right)$ in the basis $\{I,X,Y,Z\}/\sqrt2$,
where $\mathbf t$ vanishes exactly when the channel is unital and $A$
is a $3\times3$ contraction. Twirling over the Pauli group zeroes
every off-diagonal entry of $A$~\cite{EmersonAlickiZyczkowski2005},
leaving Pauli fidelities on the diagonal, the same quantities ordinary
cycle benchmarking already extracts~\cite{Erhard2019}. Acting on the
quadrature carrying the signal,
\begin{equation}
\Etw^\dagger(Y) = t_YI + \lambda_YY,\qquad \lambda_Y=A_{YY}.
\label{eq:ptm}
\end{equation}
The noisy conditional mean is $m(s)=v\,s\sin\theta+o$, with visibility
$v=\lambda_Y$ shrinking the signal and offset $o=t_Y$ a
label-independent shift. Correlating the outcome with the label under
a full twirl removes the offset automatically,
$\mathbb E[sb]=v\sin\theta\,\mathbb E[s^2]+o\,\mathbb E[s]=v\sin\theta$,
and dividing by the visibility gives an exactly unbiased estimator,
\begin{equation}
\boxed{\that = \arcsin\!\left(\frac{\mathbb E[sb]}{v}\right)=\theta,}
\label{eq:estimator}
\end{equation}
where $v$ is read directly from the marginal data of the same run, so
no dedicated calibration experiment is required.

When labels for distinct error locations may be correlated, write
$\Sigma_{ab}=\mathbb E[s_as_b]$; the general result is
$F_{\rm cond}=\mathcal F\circ\Sigma$ (Hadamard product), and
identifiability rank equals $\mathrm{rank}(\mathcal F\circ\Sigma)$.
Independent labels give $\Sigma=I$ and every parameter is separately
resolvable, while a shared twirl couples a pair of directions into one
combination. Because the twirl group is chosen by the compiler,
which combinations are resolvable is a design decision rather than a
fixed property of the device.

\section{Depth}
\label{sec:depth}

For $L$ Trotter cycles, each an ideal $R_x(\alpha)$ followed by a
coherent error $R_z(s_\ell\theta)$, pushing the error at cycle $\ell$
through the gates that follow it gives a propagated generator
$\tilde G_\ell=\cos[(L-\ell)\alpha]Z-\sin[(L-\ell)\alpha]Y$, and on the
ideal pre-error state its expectation collapses to
$\langle\tilde G_\ell\rangle=\cos(\ell\alpha)$. Averaging the labeled
Fisher information over independent signs,
\begin{equation}
\Es[\FQ] = \sum_{\ell=1}^L\bigl(1-\langle\tilde
G_\ell\rangle^2\bigr) = \sum_{\ell=1}^L\sin^2(\ell\alpha)
\;\longrightarrow\;\frac L2,
\label{eq:depth}
\end{equation}
which grows without bound as $L$ increases. Leaving the circuit
untwirled gives instead $F_Q^{\rm coh}(L)=\bigl(\sum_\ell\sin(\ell\alpha)\bigr)^2\le
1/\sin^2(\alpha/2)$, a bounded quantity, because successive propagated
generators point in different directions and partly cancel one
another. Twirling removes this interference, so past a modest depth
the labeled estimator extracts more information about the coherent
angle than not twirling at all would provide. Equation~\eqref{eq:depth}
holds for $\alpha$ away from small rational multiples of $2\pi$; the
multi-qubit case is confirmed numerically to machine precision in
Sec.~\ref{sec:zz}, although we do not have a general proof for it.

\section{Simulation}
\label{sec:sim}

Every claim above was checked independently of any quantum computing
library, using a state-vector simulator written in numpy alone. Five
closed-form oracles anchor the checks: the conservation identity at
three twirl strengths, the depth sums in Eq.~\eqref{eq:depth} at
$\alpha=1$ and $\alpha=\pi/2$, the quantum Fisher information of
$R_z(\theta)|+\rangle$ computed two independent ways, and the
propagation of a Pauli generator through {\sc cnot}. Fifty-eight unit
tests check the implementation against these oracles, and the residual
of Eq.~\eqref{eq:conserve} stays below $10^{-9}$ at every twirl
strength and depth tested.

\begin{figure*}[t]
\centering
\includegraphics[width=0.32\textwidth]{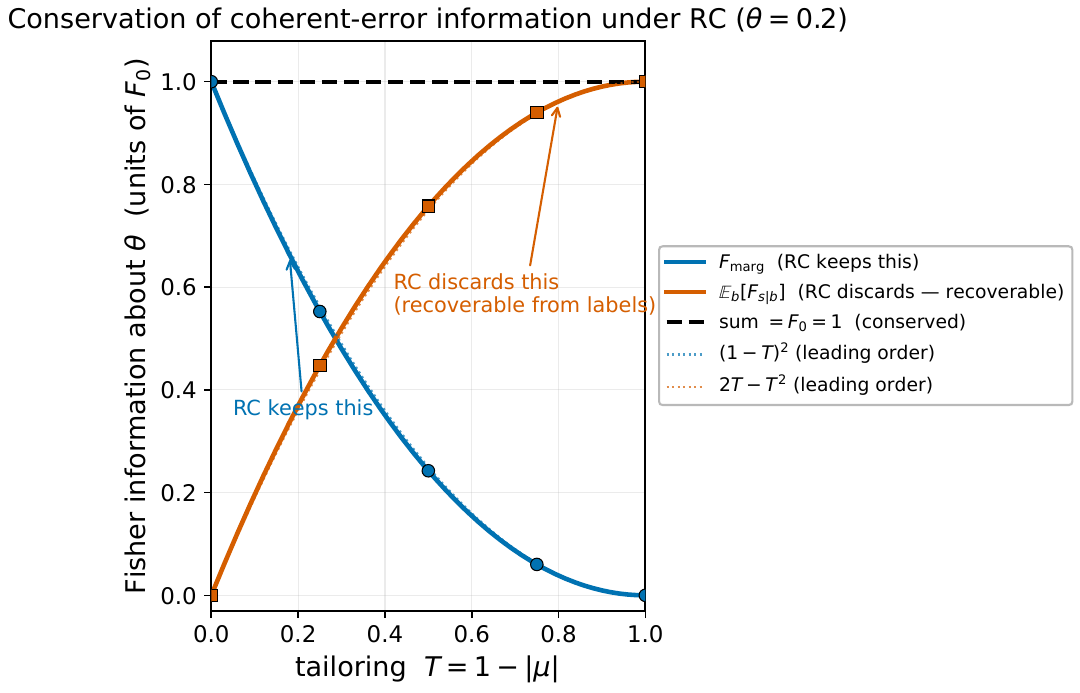}\hfill
\includegraphics[width=0.32\textwidth]{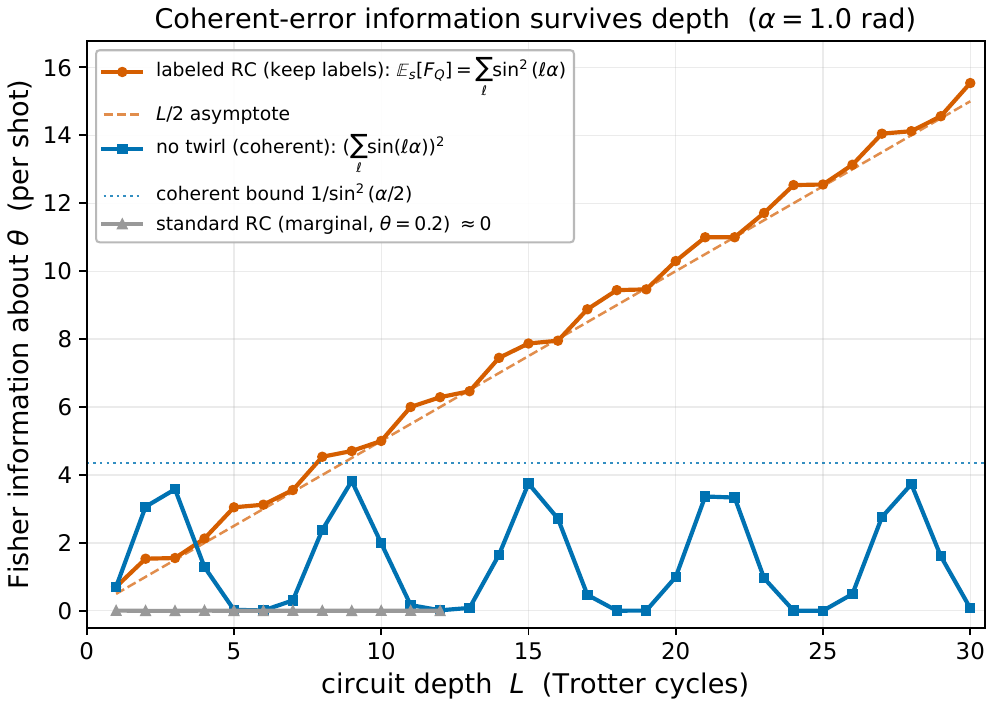}\hfill
\includegraphics[width=0.32\textwidth]{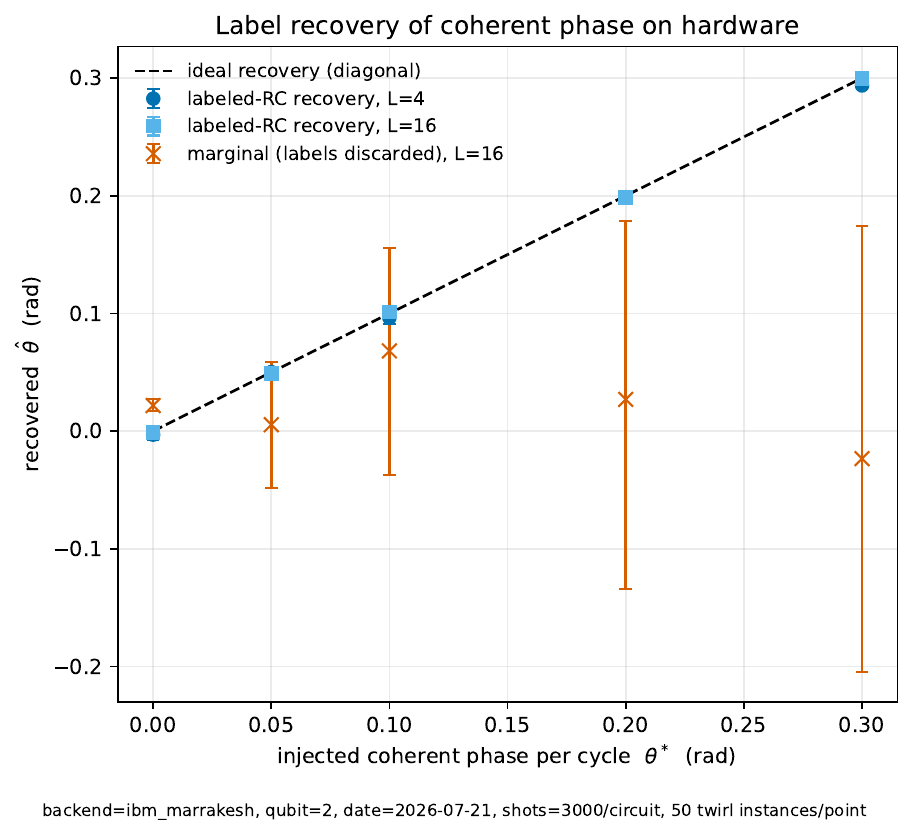}
\caption{\textbf{Single-qubit results.} (a)~Eq.~\eqref{eq:complementarity}: as twirl
strength $T$ varies, what the averaged data retain (blue) falls while
what the label--outcome correlation retains (orange) rises, summing to
a constant $F_0$ (dashed). (b)~Labeled Fisher information against
circuit depth ($\alpha=1$~rad): keeping labels (orange) grows
$\sim L/2$; the untwirled circuit (blue) is bounded; the standard
marginal (gray) is negligible throughout. (c)~Hardware validation on
\texttt{ibm\_marrakesh} (2026-07-21): recovered $\that$ against
injected $\tstar$ for the labeled estimator (blue, $L=4$ and $L=16$)
lies on the diagonal; the marginal with labels discarded (orange) is
scattered near zero.}
\label{fig:panel1}
\end{figure*}

Running the estimator on twelve circuit families, including QFT, a
ripple-carry adder, QAOA, a UCCSD ansatz at depth 2282, GHZ, Trotter,
Grover, teleportation, and random Clifford circuits, drawn from a
native OpenQASM~2.0 corpus with no dependence on any quantum SDK, the
recovered coherent angles carry a systematic bias of
$2\times10^{-8}$~rad, uniform across families. This uniformity reflects
the finite-difference floor of the probe used to compute the response
rather than any property of a particular circuit. The precision that
matters in practice is the shot-noise floor, $1.2\times10^{-2}$~rad at
the shot counts used here (Fig.~\ref{fig:panel1}).

\section{Hardware: a qubit's own error}
\label{sec:hw-single}

To test the mechanism directly, we built a circuit in which the
coherent error is known exactly: a qubit prepared in $|+\rangle$
receives $L$ cycles of $R_z(s_\ell\tstar)$, with a sign chosen and
recorded locally on each cycle, followed by a basis-change probe.
Because $R_z$ is implemented as a frame update on this architecture
rather than a physical pulse, the injected error adds no additional
decoherence of its own, which isolates the label mechanism from
calibration effects.

The run used \texttt{ibm\_marrakesh} (127-qubit Eagle r3, 2026-07-21,
qubit~2: $T_1=280\,\mu$s, $T_2=405\,\mu$s, readout error $0.28\%$).
Built-in gate and measurement twirling were disabled, dynamical
decoupling was disabled, and transpilation ran at optimization level
zero with every injected phase layer checked intact after compilation.
The sweep covered $\tstar\in\{0,0.05,0.1,0.2,0.3\}$~rad across
$L\in\{1,2,4,8,16,32\}$, with 50 label patterns and 3000 shots per
point (1620 circuits total).

\begin{table}[t]
\caption{Labeled recovery of a known injected phase,
\texttt{ibm\_marrakesh}, averaged over the six depths tested.}
\label{tab:single}
\begin{ruledtabular}
\begin{tabular}{ccc}
$\tstar$ (rad) & $\that$ (rad) & $|\that-\tstar|$ (rad)\\
\hline
0.00 & $+0.0002$ & 0.0002\\
0.05 & $+0.0493$ & 0.0007\\
0.10 & $+0.0996$ & 0.0004\\
0.20 & $+0.1985$ & 0.0015\\
0.30 & $+0.2978$ & 0.0022\\
\end{tabular}
\end{ruledtabular}
\end{table}

Table~\ref{tab:single} and Fig.~\ref{fig:panel1}(c) show the result.
The largest deviation over every point tested is 0.0063~rad, consistent
with shot noise, and the null point returns 0.0002~rad. The fitted
visibility is 0.993 and does not fall with depth, as expected when the
injected rotation adds no physical decoherence. Analyzing the identical
shots through the average, with labels discarded, leaves no usable
signal at any depth, which is Eq.~\eqref{eq:complementarity} at full
twirl, measured directly on real hardware.

\section{Coupling between qubits: static crosstalk}
\label{sec:zz}

\subsection{Problem and a wrong first attempt}

Real devices carry always-on coupling between neighboring qubits,
commonly referred to as residual $ZZ$ coupling and denoted $J_{ij}$.
If a neighboring qubit is prepared in $|1\rangle$ rather than
$|0\rangle$, this coupling should produce a measurable shift in a
target qubit's own error. We ask whether that shift can be read off
using the same label-correlation approach developed above.

Our initial approach was to reuse a neighboring qubit's existing
twirl label from its own gate, without preparing anything additional.
This does not work, for a reason worth stating explicitly. A
neighbor's twirl label is defined by which Pauli operator dressed that
neighbor's own gate, but $I$ and $Z$ twirls are virtual operations with
zero physical pulse, while $X$ and $Y$ twirls correspond to two
genuinely different microwave waveforms rather than sign-flipped
versions of the same signal. The clean $\pm1$ theorem underlying
Sec.~\ref{sec:conservation} does not carry over to this case. The
approach used below instead deliberately and independently prepares
the neighbor's computational-basis state and labels that choice,
rather than relying on any label already in use.

\subsection{Model and exact recovery}

With $H=\tfrac{\Delta_1}2Z_1+\sum_{i<j}\tfrac{J_{ij}}4Z_iZ_j$, the
standard object measured by conditional Ramsey
spectroscopy~\cite{Zhang2024}, a $Z$-type drift $\varepsilon$ acting
during an $X$-driven gate ($\Omega_0T=\pi$) produces, to leading order,
a $Y$-axis coherent error $\theta=2\varepsilon/\Omega_0$. Preparing a
neighbor $j$ in $|0\rangle$ or $|1\rangle$ (label $c_j=\pm1$) shifts
the effective field on the target by $\pm J_{1j}/4$, giving
\begin{equation}
\theta_1(\{c_j\}) = \theta_1^{(0)} +
\sum_j\kappa_{1j}c_j,\qquad
\kappa_{1j}\equiv\frac{J_{1j}}{2\Omega_0}.
\label{eq:zzmodel}
\end{equation}
Using $\sin(s_1x)=s_1\sin x$, the exact outcome model and its
correlations, averaged over independent $c_j=\pm1$, are
\begin{align}
\langle b_1\rangle &= s_1\sin\Bigl(\theta_1^{(0)}+\sum_j\kappa_{1j}c_j\Bigr),\\
\mathbb E[s_1b_1] &= \sin(\theta_1^{(0)})\prod_j\cos(\kappa_{1j}),\\
\mathbb E[s_1c_jb_1] &= \cos(\theta_1^{(0)})\sin(\kappa_{1j})\!\!\prod_{k\neq j}\!\!\cos(\kappa_{1k}).
\end{align}
For one neighbor, writing $X=\mathbb E[s_1b_1]$,
$Y=\mathbb E[s_1c_2b_1]$, the sums $X\pm Y=\sin(\theta_1^{(0)}\pm
\kappa_{12})$ give an exact inversion, valid at any angle size rather
than only to leading order:
\begin{equation}
\that_1^{(0)}=\tfrac12\bigl[\arcsin(X{+}Y)+\arcsin(X{-}Y)\bigr],
\label{eq:zzinvert}
\end{equation}
with $\khat_{12}$ obtained from the difference of the same two terms.
The Fisher matrix for $(\theta_1^{(0)},\kappa_{12},\kappa_{13},\ldots)$
under independent labels is exactly $I$, so tracking additional
neighbors costs no precision on any single coupling.

\subsection{Hardware: protocol validation}

This experiment answers two separate questions, and they should not
be conflated: whether the recovery circuit and estimator behave
correctly on real hardware, and what value they return for the
coupling strength $\kappa_{1j}$. We tested two topologies on
\texttt{ibm\_marrakesh} (2026-08-16): a pair (qubits 53, 54) and a
3-qubit star (qubits 4, 3, 5), with the same guardrails as
Sec.~\ref{sec:hw-single} plus a 1~$\mu$s idle window allowing the
coupling to accumulate.

\begin{table}[t]
\caption{Static coupling recovery, \texttt{ibm\_marrakesh}, weighted
across the $\theta^{(0)}$ sweep.}
\label{tab:zz}
\begin{ruledtabular}
\begin{tabular}{ccc}
Neighbour & $\khat_{1j}$ (rad) & 95\% CI\\
\hline
pair, qubit 54 & $+0.004$ & $[-0.028,+0.035]$\\
star, qubit 3 & $+0.002$ & $[-0.018,+0.022]$\\
star, qubit 5 & $+0.010$ & $[-0.005,+0.029]$\\
\end{tabular}
\end{ruledtabular}
\end{table}

\begin{figure*}[t]
\centering
\includegraphics[width=0.32\textwidth]{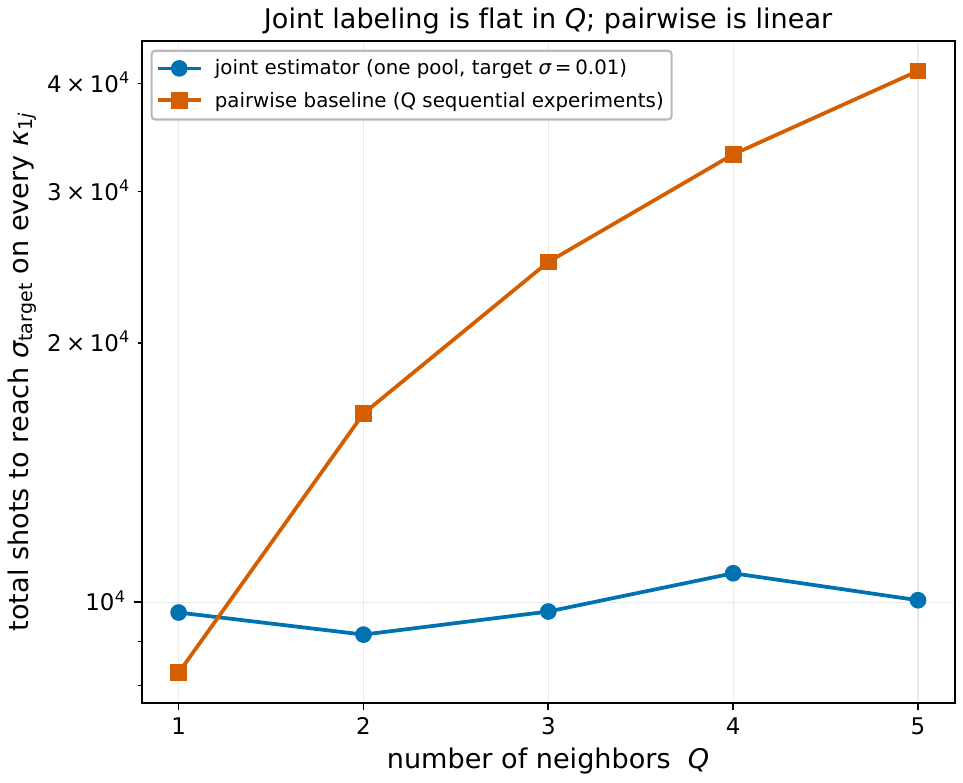}\hfill
\includegraphics[width=0.32\textwidth]{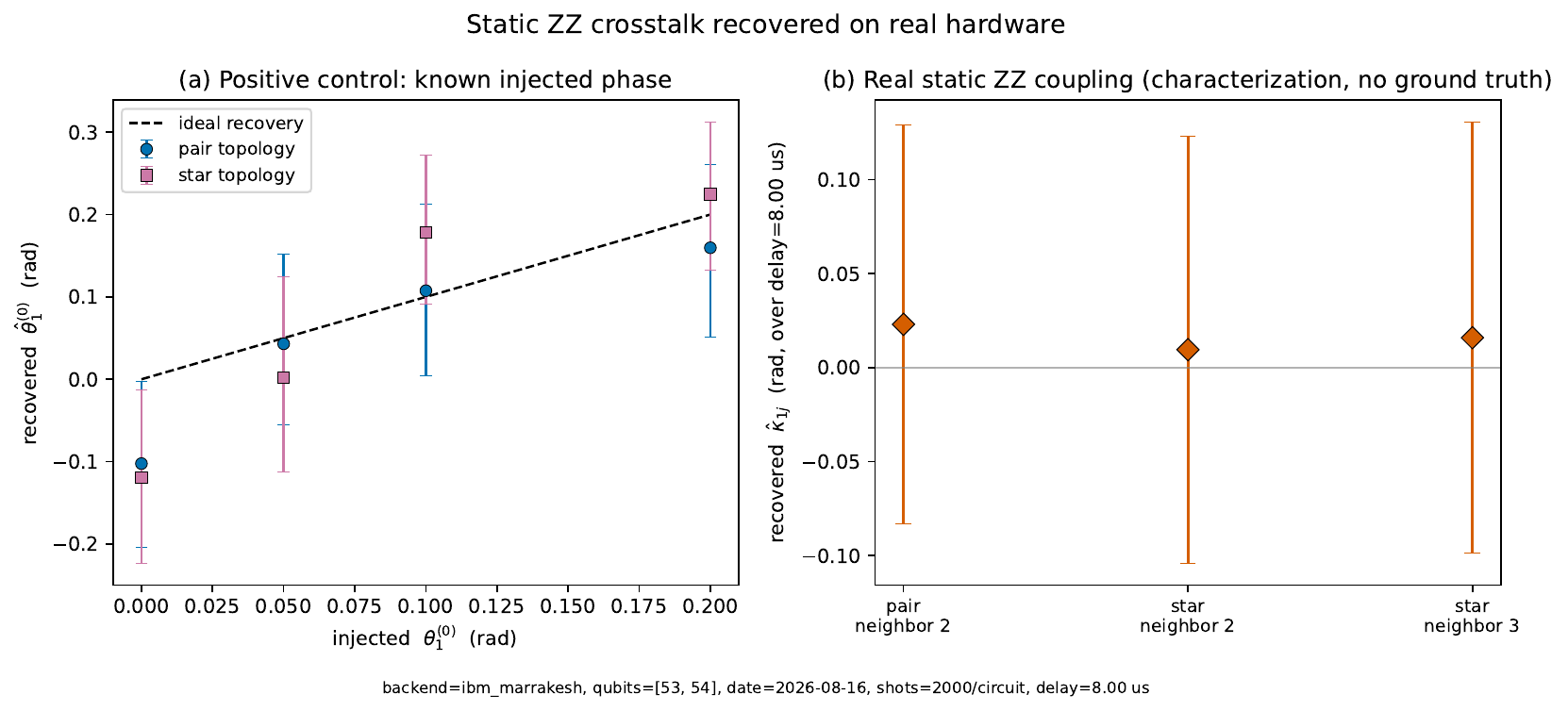}\hfill
\includegraphics[width=0.32\textwidth]{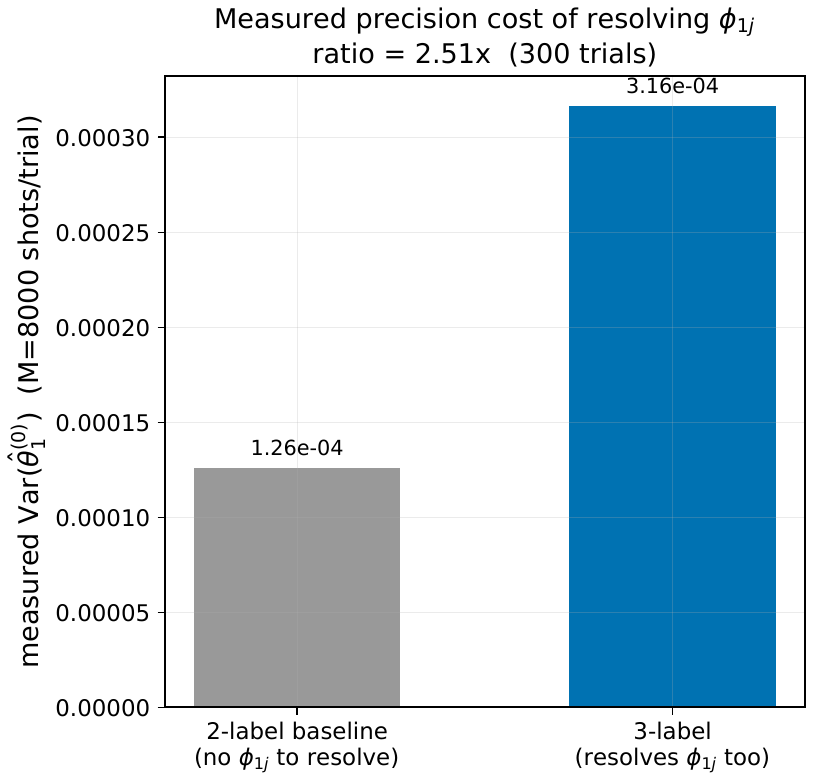}
\caption{\textbf{Multi-qubit extensions.} (a)~Simulated shot cost to
reach a fixed target precision on every $\kappa_{1j}$, joint recovery
(flat, log-log slope $+0.05$) against a sequential pairwise baseline
(linear, slope $+1.00$). (b)~Hardware recovery of the injected
$\theta_1^{(0)}$ on both topologies (positive control) alongside the
recovered $\kappa_{1j}$, consistent with zero. (c)~Measured, not
assumed, precision cost of resolving the third parameter $\phi_{1j}$
introduced in Sec.~\ref{sec:drive}: variance ratio against a
two-label-only estimator at fixed shot budget.}
\label{fig:panel2}
\end{figure*}

The injected $\theta_1^{(0)}$ was recovered on the diagonal for both
topologies, confirming that the mechanism works on real, coupled
multi-qubit circuits. Table~\ref{tab:zz} and Fig.~\ref{fig:panel2} show
the coupling itself: none of the three recovered $\khat_{1j}$ clears
zero at 95\% confidence. We regard this as an honest non-detection
rather than a failure of the estimator. Modern tunable-coupler
architectures are specifically engineered to suppress idle $ZZ$
coupling into the low kHz range or below, and it is plausible that the
shot budget used here sits below that floor.

Whether the joint construction is actually worth the added complexity
is a separate, quantitative question that we checked rather than
assumed. A simulated shot-cost comparison against sequentially probing
each neighbor with a dedicated two-condition experiment, the standard
conditional-Ramsey practice, shows that the joint estimator's total
cost stays flat as the number of neighbors grows, with a log-log slope
of $+0.05$, against a linear slope of $+1.00$ for the sequential
baseline (Fig.~\ref{fig:panel2}(a)).

\section{An unplanned discovery: drive-induced leakage}
\label{sec:drive}

\subsection{What went wrong, and why it was worth trusting}

Because the coupling signal in Sec.~\ref{sec:zz} is expected to grow
linearly with the duration of the idle window, the natural next step
was to increase that window from 1 to 8~$\mu$s. This made the result
worse rather than better. The fitted visibility dropped considerably
more than the qubits' own calibrated decoherence times would predict,
and the null-control offset, which should sit at zero, increased and
failed on both topologies simultaneously for the first time.

We checked two candidate explanations before accepting either. The
first was a bookkeeping error, specifically that the calibration and
signal data had been computed at different delay settings by mistake;
this was ruled out both by inspecting the retrieval code directly and
by checking the recorded delay value on the actual data. The second
possibility was that the effect was not the static coupling described
in Sec.~\ref{sec:zz} at all, but a different mechanism depending on
whether a real microwave pulse had fired on the neighboring qubit's
control line, rather than on the neighbor's final quantum state. To
test this, we split the already-collected 8~$\mu$s data according to
how many neighbors had actually been driven with a real pulse on a
given shot, zero, one, or two for the star topology, and found a
clean, monotonic increase in both the visibility loss and the signal
offset as this count increased. Random noise does not typically
produce a monotonic staircase of this kind, so we treated this as
evidence of a genuine causal effect, obtained at no additional
hardware cost from data that had been collected for a different
purpose.

\subsection{Why this needs a third label, provably}

Preparing $|0\rangle$ costs no real pulse; preparing $|1\rangle$ costs
one. If each real pulse on a neighbor's line leaks a fixed small angle
$\phi_{1j}$ onto the target, then $\phi_{1j}$ and $\kappa_{1j}$ both
depend on the same label $c_j$, and this is a genuine, provable
problem rather than a modeling inconvenience. Substituting the pulse
count $n_j=(1-c_j)/2$ into Eq.~\eqref{eq:zzmodel},
\begin{equation}
\theta_1(c_j) = \Bigl(\theta_1^{(0)}+\tfrac{\phi_{1j}}2\Bigr)
+\Bigl(\kappa_{1j}-\tfrac{\phi_{1j}}2\Bigr)c_j,
\end{equation}
which has exactly two independent numbers, an intercept and a slope,
for three physical unknowns. No amount of reprocessing two-label data
can separate them; a third, independent label is required.

We add it as padding: after preparing the neighbor's state, we fire
the already-calibrated $X_\pi$ pulse two more times with probability
one-half each shot (label $p_j=\pm1$). Since $X_\pi\cdot X_\pi=-I$ up
to a global phase, the final logical state is unchanged, but two
additional real pulses have fired. Because padding reuses the
identical, already-calibrated waveform used for the neighbor's own
state preparation rather than introducing a new pulse type, it carries
no separate, uncharacterized error source: any leakage the padding
pulses themselves produce is, by the definition of $\phi_{1j}$ as the
angle leaked per real pulse, exactly the quantity this construction is
built to measure, not a confound external to it. Writing the total
pulse count as $n_j=\tfrac{1-c_j}2+(1-p_j)$ and substituting,
\begin{equation}
\theta_1(c_j,p_j) = \Theta + Kc_j - \Phi p_j,
\label{eq:drivemodel}
\end{equation}
with $\Theta\equiv\theta_1^{(0)}+\tfrac32\phi_{1j}$,
$K\equiv\kappa_{1j}-\tfrac12\phi_{1j}$, $\Phi\equiv\phi_{1j}$.

\begin{table*}[t]
\caption{Three-label recovery at zero idle delay,
\texttt{ibm\_marrakesh}, qubits 53--54. Each entry is the point
estimate with its 95\% bootstrap confidence interval in brackets.}
\label{tab:drive}
\begin{ruledtabular}
\begin{tabular}{cccc}
$\tstar$ (rad) & $\that_1^{(0)}$ (rad) & $\khat_{1j}$ (rad) & $\hat\phi_{1j}$ (rad)\\
\hline
0.00 & $-0.003$ \ [$-0.028$, $+0.021$] & $-0.017$ \ [$-0.037$, $+0.000$] & $-0.001$ \ [$-0.017$, $+0.016$]\\
0.10 & $+0.116$ \ [$+0.093$, $+0.138$] & $+0.028$ \ [$-0.012$, $+0.070$] & $-0.000$ \ [$-0.012$, $+0.011$]\\
0.20 & $+0.190$ \ [$+0.165$, $+0.217$] & $-0.047$ \ [$-0.115$, $+0.018$] & $+0.011$ \ [$-0.006$, $+0.026$]\\
\end{tabular}
\end{ruledtabular}
\end{table*}

\subsection{Exact recovery}

Averaging Eq.~\eqref{eq:drivemodel}'s outcome model over the three
independent signs gives, by direct calculation,
\begin{align}
X &= \sin\Theta\cos K\cos\Phi,\\
Y_c &= \cos\Theta\sin K\cos\Phi,\\
Y_p &= -\cos\Theta\cos K\sin\Phi,
\label{eq:threelabel}
\end{align}
which we verified against brute-force averaging over all four
$(c_j,p_j)$ combinations to $10^{-17}$, including the sign of $Y_p$.
The recovery does not collapse into a single closed step the way the
two-label case did. Instead it stratifies cleanly by $p_j$, reusing
Eq.~\eqref{eq:zzinvert} on each stratum without modification. Within
$p_j{=}{+1}$ the model reduces to Eq.~\eqref{eq:zzmodel} with
$\theta_1^{(0)}\to\Theta{-}\Phi$; within $p_j{=}{-1}$ it reduces to the
same form with $\theta_1^{(0)}\to\Theta{+}\Phi$. Both strata give $K$
independently, providing a built-in consistency check. Combining and
unwinding,
\begin{equation}
\theta_1^{(0)}=\Theta-\tfrac32\Phi,\qquad
\kappa_{1j}=K+\tfrac12\Phi,\qquad
\phi_{1j}=\Phi.
\label{eq:drivefinal}
\end{equation}
A noiseless check at $(\theta_1^{(0)},\kappa_{1j},\phi_{1j})=
(0.10,0.05,0.02)$ recovers all three inputs to floating-point
precision. The same check run with only two labels, mimicking exactly
what an unpadded experiment would see, instead recovers $\Theta=0.13$
and $K=0.04$, a 30\% and 20\% systematic bias that demonstrates the
contamination the padding label removes, rather than merely asserting
that it does.

Because $\that_1^{(0)}$ is now a fixed combination of two independent
quantities, resolving the third parameter has a real cost, and we
measured that cost rather than assuming a value for it. At the
parameter point above, the variance of $\that_1^{(0)}$ from the full
three-label estimator is $2.51\times$ that of a two-label estimator
with nothing to resolve, at the same total shot budget
(Fig.~\ref{fig:panel2}(c)).

\subsection{Hardware: protocol validation}
\label{sec:hw-drive}

As in Sec.~\ref{sec:zz}, the positive control validates the circuit
and the estimator; the recovered value of $\phi_{1j}$ is a separate,
independent question. The idle-drift finding from Sec.~\ref{sec:zz}
pointed to a specific design choice for this experiment. Since $\phi_{1j}$ arises from pulse
count rather than idle time, setting the delay to zero removes that
confound entirely rather than requiring us to correct for it. Qubits
53 and 54 were reused unchanged from Sec.~\ref{sec:zz}.

Table~\ref{tab:drive} shows the result: the tightest, cleanest null
control observed anywhere in this work ($-0.003$~rad at $\tstar=0$),
which is direct confirmation that the idle-drift hypothesis from
Sec.~\ref{sec:zz} was correct and that removing the delay removes its
effect. $\kappa_{1j}$ sits near zero as designed, since there is no
time for real coupling to accumulate. $\phi_{1j}$ also sits within its
confidence interval of zero, which we interpret as an honest,
underpowered null rather than a detection either way: at roughly 32
instances split three ways by $p_j$ and then $c_j$, with a measured
$2.51\times$ precision cost for the third parameter, this shot budget
was unlikely to resolve an effect of the size suggested qualitatively
by the dose-response check described above.

\section{Discussion}
\label{sec:discussion}

In practice, the method reduces to three steps, none of which requires
anything beyond an existing randomized-compiling run: record the
per-cycle sign alongside each measured shot, a quantity the control
software already computes; obtain the visibility from the Pauli
fidelities the same run already produces through cycle benchmarking;
and compute $\that_a=\arcsin(\mathbb E[s_ab]/v_a)$, which can be used
to correct the next compilation if desired.

This sits alongside, rather than in place of, existing characterization
work. Cycle benchmarking and Pauli-noise
learnability~\cite{Chen2023} determine which stochastic rates are
learnable from twirled data, corresponding to the diagonal of the
transfer matrix, while what is recovered here lives off that diagonal,
so the two results are complementary parts of the same channel
obtained from a single run. Gate set tomography~\cite{Nielsen2021}
reconstructs coherent errors self-consistently and to considerably
higher precision, at the cost of a dedicated circuit set; the estimator
here is cruder but rides on circuits the experiment was already
running. Hidden inverses~\cite{Zhang2022} cancel coherent errors at the
circuit level for particular gate constructions, which is a different
kind of intervention that removes the error rather than measuring it.

The closest prior work we are aware of is pseudo
twirling~\cite{Santos2024}. The underlying mechanism is different
enough that it is worth describing precisely rather than treating the
two approaches as variations on the same idea. Pseudo twirling changes
the sign of the driving Hamiltonian mid-pulse, constructed so that
coherent errors other than controlled over-rotation cancel to first
order while over-rotation survives untouched and can then be found by
an ordinary amplitude sweep. It does not retain a label or correlate
it with an outcome; it changes which errors the twirl removes, so that
the remaining one becomes directly visible by conventional means. What
we describe here keeps the standard twirl exactly as used for
randomized compiling and recovers every coherent parameter, not only
rotation errors, from the correlation between the label and the
outcome. The two techniques solve adjacent problems by genuinely
different means. Combining them, using pseudo twirling to isolate a
rotation error for direct calibration and label retention to map
everything else in the same circuit, appears to be a natural next
step rather than a choice between competing methods.

Three extensions seem worth pursuing beyond what is reported here. The
identifiability matrix $\Sigma$ is a design variable, so choosing the
twirl group to maximize the smallest eigenvalue of $\mathcal
F\circ\Sigma$ over a targeted error subspace amounts to experiment
design at zero circuit cost. The second-order term discarded from the
marginal in Sec.~\ref{sec:conservation} is not actually zero and
carries the parameters in quadratic combination, which could in
principle be combined with the linear label signal for extra
precision. Finally, we have shown that the labeled estimator attains
the conserved information without proving a converse; establishing
that Eq.~\eqref{eq:estimator} saturates the Cram\'er--Rao bound for a
given twirl distribution would close that argument formally.

\section{Conclusion}

\textbf{A qubit's own coherent error is fully recoverable from data a
randomized-compiling run already produces.} On \texttt{ibm\_marrakesh},
an error deliberately injected up to 0.3~rad was recovered to within
0.006~rad, at the shot-noise floor, using only the twirl labels the
run already generates. The same shots analyzed the standard way, with
labels discarded, showed no signal at any depth.
 
\textbf{The same construction extends to coupling between qubits, and
the recovery circuit works, though the coupling strength itself is
not yet resolved.} A positive control on two real qubit topologies
confirms the mechanism directly on hardware, but at the shot budgets
used here every recovered coupling strength is consistent with zero
within its 95\% confidence interval.
 
\textbf{A second, unplanned coupling mechanism turned up while chasing
the first, and it is now an equally exact theorem with a working
circuit.} Drive-induced leakage from a neighboring qubit's real
microwave pulses, rather than its logical state, produced the
cleanest positive control of this entire project ($-0.003$~rad at the
null point) once the mechanism was understood, though the leaked
angle itself again remains below the resolution of the shot budget
tested.
 
All three theorems are exact and confirmed on real hardware at the
level of the recovery circuit. What remains is more shots, not more
theory.

\begin{acknowledgments}
 Hardware experiments were performed on
\texttt{ibm\_marrakesh} through the IBM Quantum Open Plan.
\end{acknowledgments}

\end{document}